\documentclass[12pt]{article}
\usepackage{epsfig}
\usepackage{amssymb, amsmath}
\newlength{\abstractwidth}
\newcommand{\starttext}{
\setcounter{footnote}{0}
\renewcommand{\thefootnote}{\arabic{footnote}}}
\newcommand{\be}{\begin{equation}}
\newcommand{\ee}{\end{equation}}

\def\ra{\rightarrow}

\def\IC{\mathbb{C}}
\def\IR{\mathbb{R}}
\def\IZ{\mathbb{Z}}

\def\TU{\tilde U}
\def\TV{\tilde V}
\def\TO{\tilde 1}
\def\TT{\tilde 2}
\def\ICC{\IC}
\def\ID{\relax{\rm I\kern-.18em D}}
\def\IE{\relax{\rm I\kern-.18em E}}
\def\IF{\relax{\rm I\kern-.18em F}}
\def\IG{\relax\hbox{$\inbar\kern-.3em{\rm G}$}}
\def\IGa{\relax\hbox{${\rm I}\kern-.18em\Gamma$}}
\def\IH{\relax{\rm I\kern-.18em H}}
\def\II{\relax{\rm I\kern-.18em I}}
\def\IK{\relax{\rm I\kern-.18em K}}
\def\IP{\relax{\rm I\kern-.18em P}}
\def\I1{{\bf 1}}
\def\ZNZN{\ICC^3 / \ \! \IZ_N \times \IZ_N }
\def\ZN2{\IZ_N \times \IZ_N}

\def\IR{\relax{\rm I\kern-.18em R}}

\def\OL#1{ \kern1pt\overline{\kern-1pt#1\kern-1pt}\kern1pt }

\begin{document}
\renewcommand{\theequation}{\thesection.\arabic{equation}}

\begin{titlepage}
\bigskip
\rightline{SU-ITP 01/46}
\rightline{hep-th/0111079}

\bigskip\bigskip\bigskip\bigskip

\centerline{\Large \bf {Deconstructing Noncommutativity}}
\bigskip
\centerline{\Large \bf {with a}}
\bigskip
\centerline{\Large \bf {Giant Fuzzy Moose}}

\bigskip\bigskip
\bigskip\bigskip

\centerline{Allan Adams\footnote{allan@slac.stanford.edu} and
Michal Fabinger\footnote{fabinger@itp.stanford.edu}}
\medskip\medskip\medskip\medskip
\centerline{Department of Physics and SLAC}
\centerline{Stanford University}
\centerline{Stanford, CA 94305-4060}

\medskip
\medskip
\bigskip\bigskip
\begin{abstract}

We argue that the worldvolume theories of D-branes probing orbifolds with discrete torsion
develop, in the large quiver limit, new non-commutative directions.
This provides an explicit `deconstruction' of a wide class of
noncommutative theories.
This also provides insight into the physical
meaning of discrete torsion and its relation to the T-dual B field.
We demonstrate that the strict large quiver limit reproduces the
matrix theory construction of higher-dimensional D-branes,
and argue that finite `fuzzy moose' theories provide novel
regularizations of non-commutative theories and explicit
string theory realizations of gauge theories on fuzzy tori.
We also comment briefly on the relation to $NCOS$, $(2,0)$
and little string theories.

\medskip
\noindent
\end{abstract}

\end{titlepage}
\starttext \baselineskip=18pt
\setcounter{footnote}{0}
\setcounter{equation}{0}

\section{Introduction}
Recent work on the phenomenology of large $N$ gauge theories has revealed that
theories based on `moose diagrams' (aka `quiver theories')
generate extra dimensions in the large quiver limit.
Explicitly, the lagrangian takes the form of a lattice field theory plus extra irrelevant
matter that completes the theory in the UV.
This so-called `dimensional
deconstruction'\cite{Deconstruction} (see also \cite{Hill},\cite{Phenomenology})
has an elegant realization in string theory, where the quiver theories arise
as the worldvolume theories of D-branes on geometric orbifolds.
Some rather clever moose phonomenology
has led to novel presentations of several interesting systems,
including the mysterious interacting $6d$ $(2,0)$  and little
string theories\cite{DeconstructingLS}, hinting that there is more
to be learned from this subtle regulation scheme.

Of course, there already exist methods for generating higher-dimensional branes,
in particular matrix theory constructions\cite{MatrixDBranes, Seiberg, Myers,
RaamsdonkTaylor} which lead to non-commutative worldvolumes.
What is the relation between these two approaches?
Can one deconstruct noncommutative dimensions?

In this note we present evidence that the worldvolume theories of D-branes probing
$\IC^3/\IZ_N\times\IZ_N$ orbifolds with discrete torsion provide, in the
$N\to \infty$
large-moose limit, precisely such a construction.
The new non-commutative directions appear exactly as in the original case;
their non-commutativity derives directly from the
extra phases specifying the discrete torsion.

This can be motivated as follows. First, as we will explain in
detail below, the partition function for closed strings on the
torus $\IR^2 /\IZ\times\IZ$ with discrete torsion
$\epsilon=e^{2\pi i /n}$, $1/n \in (0,1)$, is identical to that on
the same torus but with a constant background $B$-field such that 
$\epsilon=e^{2\pi i \int B}$.
Thus the theory of a D$0$-brane on a torus with discrete torsion
$\epsilon$ is dual to the theory of a D$2$-brane wrapping the
T-dual torus with a (rescaled) constant background ${B}$-field
${b} \sim 1/n$ - which is exactly noncommutative SYM on the same
torus.

Now consider the orbifold $\IC^3/\ZN2$ with discrete
torsion $\epsilon$. This is a cone over base 
$S^5/\ZN2$; for large $N$ this is a very sharp cone.
Consider a thin region of this cone far from
the orbifold fixed point, where the geometry is approximately
$\IR^4\times T^2$;
in the $N\to\infty$ limit, the local physics
precisely reproduces the toroidal orbifold with
discrete torsion.
This
implies that the
quiver theory of a D$p$-brane on $\IC^3/\ZN2$ with discrete torsion reproduces,
in the large $N$ limit, the noncommutative theory of a D$(p+2)$-brane wrapping a
torus with a constant background $B$-field specified by the discrete torsion.

In the following we explicitly verify this conjecture by
demonstrating that this quiver theory is equivalent to SYM on an
$N^2$-point fuzzy torus with a noncommutativity parameter
specified by the discrete torsion $\Theta\sim 1/b \sim
n$.\footnote{Everywhere in the paper we choose to describe NCYM in
terms of parameters given by eqs. (\ref{BAndPhi}),
(\ref{OpenVsClosed}).} (In this relationship we keep the closed
string volume fixed.) In the $N\to\infty$ limit, this precisely
reproduces SYM on a smooth torus with constant $B$-field correctly
specified by the discrete torsion. In the language of
\cite{Deconstruction}, this shows that the orbifold $\IC^3/\ZN2$
with discrete torsion can be used to ``deconstruct'' a wide class
of noncommutative theories.

In particular, one can use the $\IC^3/\ZN2$ orbifolds with
discrete torsion to deconstruct noncommutative D$p$-brane theories
for $p=2,3,4,5$. For $p=3$, one can take a further strong coupling
limit to obtain a deconstruction of NCOS theory\cite{NCOSrags}.
For $p=4,5$, the D-brane SYM theories are not UV complete; it is a
remarkable fact that the deconstructed theories appear to contain
precisely the degrees of freedom required to complete the D-brane
theories to $(2,0)$ \cite{TwoCommaZero} or little string theories
\cite{LittleStrings}, respectively, as demonstrated in
\cite{DeconstructingLS}.

We begin by reviewing the original `deconstruction' phenomenon\cite{DeconstructingLS}.
We then recall the quiver theories on branes
probing orbifolds with discrete torsion
and demonstrate how they deconstruct non-commutative theories.
We compare the deconstruction of noncommutative D-branes
to their matrix theory constructions, finding agreement in the large moose limit,
and argue that finite moose provide stringy realizations of gauge theories
on fuzzy tori.  We close with further speculations and open questions.  
(For related earlier work, see eg \cite{Dasgupta:2000hn,Berenstein:2000hy,Hanany:2000fq}.)

\section{A brief review of (de)construction}
\label{OrdinaryD-BraneDeconstruction}
Consider the worldvolume theory of a single D0-brane on a supersymmetric
$\IC ^2 / \IZ_N$ orbifold. (A D3-brane probe of this orbifold was used
in \cite{DeconstructingLS} to deconstruct the
six-dimensional (2,0) theory.) The orbifold can be thought of as a
local model for an $A_{N-1}$ singularity in a K3 manifold,
preserving half of the original
supersymmetry. A general technique for constructing worldvolume
theories of D-branes on orbifolds was described in a remarkable paper by
Douglas and Moore \cite{Quivers}; we follow their
procedure.

Parameterizing the target space with five real scalars
$x^{m}, m = 1 \ldots 5$ and two complex scalars $z^1 = x^6 + i x^7$ and $z^2 = x^8 + i x^9$, the geometric action of the
$\IZ_N$ generator is
\begin{equation}
  R(e) = \exp ( {2\pi i} (J_{67} - J_{89}) /N )\ .
\label{C2ZN-Rotation}
\end{equation}

The massless worldvolume fields of the parent ${\cal N} =4$ $U(N)$
gauge theory are gauge fields $A^0_{ij}$, scalars $X^{m}_{ij} ,\
Z^1_{ij},\ Z^2_{ij}$, and majorana-weyl spinors in the {\bf 16} of
$SO(9,1)$, $\lambda_{ij}$. As in \cite{Panic}, we  construct
$\chi$, a weyl spinor of $SO(5,1)$, out of the components of
$\lambda$ having $(s_{67}, s_{89}) = (-\frac{1}{2},- \frac{1}{2})$
or $(s_{67}, s_{89}) = ( +\frac{1}{2},+\frac{1}{2})$. Similarly
the weyl spinor $\eta$ will contain the components of $\lambda$
with either $(s_{67}, s_{89}) = (-\frac{1}{2} ,+ \frac{1}{2})$ or
$(s_{67}, s_{89}) = ( \frac{1}{2} ,- \frac{1}{2})$.

When acting on D-branes, the orbifold group may have an additional
action on the chan-paton indices. The action of the generator $e$
of the $\IZ_N$ orbifold group can thus be written
 \begin{equation}
  |\psi, i, j \rangle \ra
  \gamma(e)_{ii'}\ |R(e)\psi, i', j' \rangle \ \gamma(e)^{-1}_{j'j}
  \label{FullOrbifoldAction}
\end{equation}
where the $\gamma$-matrices belong to a faithful representation of
the orbifold group. In the case at hand, we can express $\gamma(e)$ in a
convenient basis as
\begin{equation} \gamma(e) = diag(0,e^{2 \pi i / N}, ..., e^{2 \pi i (N-1)/ N}) \end{equation}
The fields surviving the orbifold projection
(\ref{FullOrbifoldAction}) are thus
\begin{equation} A^0_{ii},\quad X^m_{ii}, \quad Z^1_{i, i+1},\quad Z^2_{i+1,i},
\quad\chi_{i,i}, \quad\eta_{i, i-1}. \end{equation}
This spectrum can be conveniently represented by so called moose or quiver
diagrams (see Fig. 1).

\smallskip
\centerline{\epsfbox{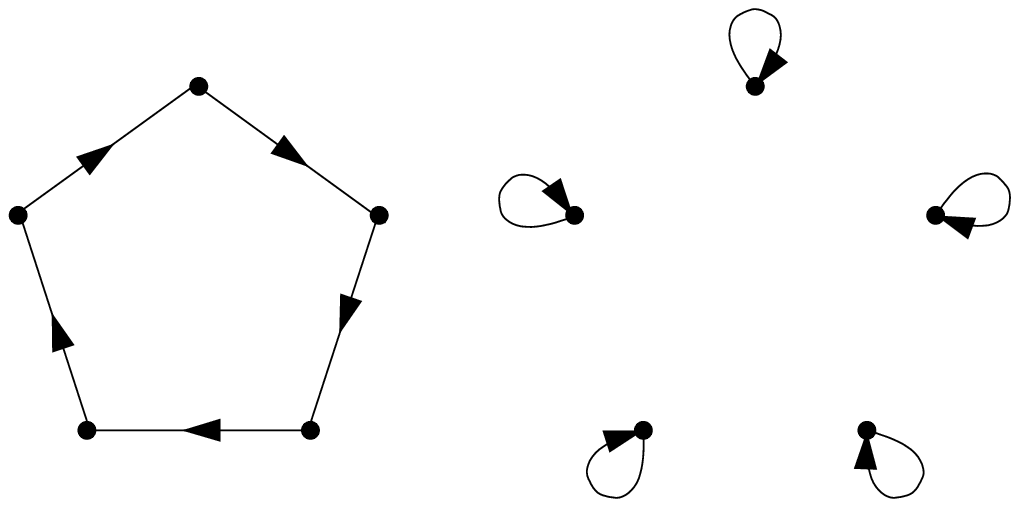}}
\smallskip
\centerline{Figure 1: The $\IC^2/\IZ_5$ Moose.
On the left is the moose for $Z^1_{i, i+1}$, ${\bar Z}^2_{i,i+1}$, and ${\bar \eta}_{i,i+1}$.}
\smallskip
\centerline{On the right is the moose for $A^0_{ii}$, $X^m_{ii}$, and $\chi_{i,i}$.}
\smallskip

The classical potential, descending from the potential of the
parent theory by restricting to fields which survive the orbifold
projection, includes the following term for every $m=1 \ldots 5$:
\begin{equation} V = \frac{1}{2gN l_s (2\pi l_s^2)^2} \sum_{j=0}^{N} \left(
X_{j+1,j+1}^m - X_{jj}^m \right)^2 \left( |Z^1_{j,j+1}|^2 +
|Z^2_{j+1,j}|^2 \right).
 \label{PotentialForX} \end{equation}
There are also  terms quartic in $Z^1$ and $Z^2$ forcing
$|Z^1_{j,j+1}|$ and $|Z^2_{j+1,j}|$ to be independent of $j$. The
moduli space has a coulomb branch, where all $Z^1_{j,j+1}$ and
$Z^2_{j+1,j}$ vanish and where $X^m_{jj}$ can be independently
varied. It also has a higgs branch with
\begin{equation}
 |Z^1_{j,j+1}| = r_1, \quad |Z^2_{j,j+1}| =
r_2. \label{HiggsZ1Z2} \end{equation}
We have added a factor of $1/|\IZ_N|=1/N$ to (\ref{PotentialForX})
so that the higgs expectation values (\ref{HiggsZ1Z2}) correspond
to moving a D0-brane (consisting of $N$ fractional D0-branes) to a
distance $r=(r_1^2+r_2^2)^{-1/2}$ from the orbifold point.

The trick is now to study low energy fluctuations around a
particular point on the higgs branch given by some fixed values of
$r_1$ and $r_2$. The leading order potential for $X_{jj}^m$, for
example, will be
\begin{equation}
 V = \frac{r^2}{2gN l_s (2\pi l_s^2)^2}
\sum_{j=0}^{N} \left( X_{j+1,j+1}^m - X_{jj}^m \right)^2.
\label{HiggsedPotentialForX} \end{equation}
For large $N$ and $X_{jj}$ slowly
varying with $j$, this looks very much like a lattice discretization of
\begin{equation}
 V = \frac{2 \pi R r^2}{gN^2 l_s (2\pi l_s^2)^2}
\int_{0}^{2\pi R}  \frac{1}{2} \left( \frac{dX^m}{d\sigma}
\right)^2 d\sigma. \label{PotentialForContinuousX} \end{equation}
with $\sigma
\in (0, 2 \pi R )$ and effective lattice spacing $a=2\pi R/N$.
It is a stimulating and life-affirming exercise to check that the
rest of the lagrangian takes the correct form to reproduce,
in the continuum limit, the
worldvolume theory of a D1-brane.

There is a simple geometric reason that this works.
For large $N$ the orbifold is a sharp cone over base $S^3/\IZ_N \sim S^2\times(S^1/\IZ_N)$.
For large $r$, far from the fixed locus,
the local geometry seen by the
D0-brane is approximately a cylinder of radius $r_c=r/N$, with the
compact coordinate being in the $J_{67} - J_{89}$ direction
\cite{DeconstructingLS,Panic}.
T-duality along this coordinate (which is valid far from the fixed locus)
produces a D1-brane wrapping an $S^1$ of radius
\begin{equation} R \equiv \tilde r_c = \frac{l_s^2}{r_c} = \frac{N l_s^2}{r}
\end{equation}
and string coupling
\begin{equation} \tilde g_s = \frac{R}{l_s} g_s = \frac{l_s}{r_c} g_s =\frac{N l_s}{r} g_s.
\end{equation}
With these values of $R$ and ${\tilde g}_s$, the factor in front of
the integral in (\ref{PotentialForContinuousX}) becomes precisely
the D1-brane tension
\begin{equation} \tau_1  = \frac{1}{\tilde g_s l_s (2\pi l_s)} . \end{equation}
It is simple and remarkable to check that the interaction terms in
the D1-brane world-volume theory thus obtained are correctly normalized.

We can of course T-dualize without
approximating the geometry by a cylinder \cite{OoguriVafa}. The T-dual
of the full orbifold geometry involves $N$
NS5-branes evenly spaced along the T-dual circle, while
the fractional D0-branes at the orbifold point become
D-strings stretched between the NS5-branes.

\subsection{Scalings}
\label{Scalings}

It is easy to find similar quiver theories which mock up higher
dimensional D-branes at energies
lower than $1/a$ (alternatively, whose IR physics in the limit of
$a \ra 0$ reproduces a higher-dimensional D-brane worldvolume
theory). In supersymmetric cases, the number of supercharges is
increased in this limit, because an orbifold generally breaks some
supersymmetry, whereas a torus with periodic boundary conditions
does not. In our example, the number of supercharges is doubled,
giving 16 real supercharges in the end.

The scalings of the various couplings also deserve attention.
Consider for concreteness the deconstruction of $5d$ $U(k)$ theory
with small but finite coupling $g_{5}^2$.  This proceeds as above
by studying light fluctuations off the higgs branch of the cyclic
moose, similar to \cite{Deconstruction}. Let $g_q$ be the gauge
coupling in the original $U(k)^N$ quiver theory. Out along the
higgs branch, the effective coupling of the surviving $U(k)$ gauge
theory is
      \begin{equation} g_4^2 = g_q^2 /N.\end{equation}
The relation between the $5d$ and $4d$ couplings is as
usual for kaluza-klein reduction,
      \begin{equation} g_5^2 = g_4^2 R.\end{equation}
Expressed in terms of the coupling in the original quiver theory,
the $5d$ coupling is thus
      \begin{equation} g_5^2 = g_q^2 R /N,\end{equation}
so holding the $5d$ coupling and (emergent) radius $R$ fixed and finite
while taking $N\to\infty$ gives finite $g_4$ but requires
taking the original coupling of the higgsed moose large,
      \begin{equation} g_q \sim N.\end{equation}
The upshot is that, while the gauge coupling of the original
quiver theory is getting large, the gauge couplings of both
the $k$ D$p$-branes far form the orbifold fixed point and
of the deconstructed D$(p+2)$-branes can be kept arbitrarily weak.

More generally, consider the deconstruction of a $(d+p)$
dimensional theory with small but finite 't hooft coupling
$kg_{p+d}^2$. In terms of the effective lattice spacing $a$ and
quiver coupling $g_q$,
\begin{equation} g^2_{p+d} \approx a^d g^2_{q}. \end{equation}
The (very strongly higgsed!) original quiver theory is thus
strongly coupled in the continuum limit $a\to 0$. On the other
hand, for $N$ large but fixed energy scale comparable to $R\sim
r/N$ in the deconstructed theory, these configurations have
extremely large higgs expectation values $r\sim N$, with the
result that the low-energy excitations are not localized in the
moose but spread over many gauge groups. This makes the effective
interactions small, so the dynamics can be studied perturbatively,
which is the statement that the effective $(p+d)$-dimensional 't
hooft coupling $kg^2_{p+d}$ can be held weak in the continuum
limit.

Second, the matrix hamiltonian, which gives for example
(\ref{PotentialForX}), is usually said to be valid only for small
separations between the D-branes. Naively, one might worry that
(\ref{PotentialForX}) becomes inapplicable when $r$ approaches the
string length $l_s$. However, as was discussed in e.g.
\cite{Myers}, the true limitation is to energies lower than the
string scale, i.e. to strings shorter than $l_s$.  This requires
\begin{equation} \frac{r}{N} \ll l_s \label{StingLengthInequality} \end{equation}
so that the model will be rich enough to accurately describe the
physics. This bound (\ref{StingLengthInequality}) translates to
requiring the lattice spacing $a$ to be {\it larger} than the
string length,
\begin{equation} a \gg l_s .\end{equation}
Happily, this is not an obstacle, as we are interested in the
decoupling limit, i.e. in the physics at energies much lower than
$1/l_s$.

\subsection{$\ZNZN$ without discrete torsion}
\label{Ordinary-C3-ZNZN-Orbifold}

Finally, let's review some of the salient properties of
the $\ZNZN$ orbifold {\it without} discrete torsion.
(This orbifold was used to deconstruct (1,1) little string
theory in \cite{DeconstructingLS}.) Choosing three real
transverse coordinates $x^1, \ x^2, \ x^3$ and three complex
coordinates $z^1=x^4 + i x^5, \ z^2=x^6 + i x^7, \ z^3 = x^8 + i
x^9$ the geometric action of the two orbifold group generators is
\begin{equation}
\begin{array}{c}
  R(e_1) = \exp( {2\pi i} ( - J_{67} + J_{89}) /N   ) \\
  R(e_2) = \exp( {2\pi i} (   J_{45} - J_{89}) /N   ),
\end{array}
\label{C3ZNZN-Rotations} \end{equation}
preserving one quarter of the original supersymmetry.
The field content corresponding to $k$ transverse type II
D$p$-branes ($p \leq 3$) descends from a configuration of
$kN^2$ D$p$-branes in the parent theory; a judicious choice
of basis gives the action on gauge indices as
\begin{equation}
\begin{array}{c}
 \gamma(e_1)_{aj \tilde k;\ a'j'\tilde k'} = (e^{2 \pi i / N})^j \
\delta_{aa'} \delta_{jj'} \delta_{\tilde k \tilde k'}  \\
\gamma(e_2)_{aj \tilde k;\ aj'\tilde k'} = (e^{2 \pi i /
N})^{\tilde k} \ \delta_{aa'}\delta_{jj'} \delta_{\tilde k \tilde
k'},
\end{array}
\label{GammaMatricesForZNZN}
 \end{equation}
where $a,a'=1\ldots k$ and $j,j',\tilde k, \tilde k'=1\ldots N$.
The surviving spectrum is summarized by the quiver diagram on Fig.
2.

\smallskip
\centerline{\epsfbox{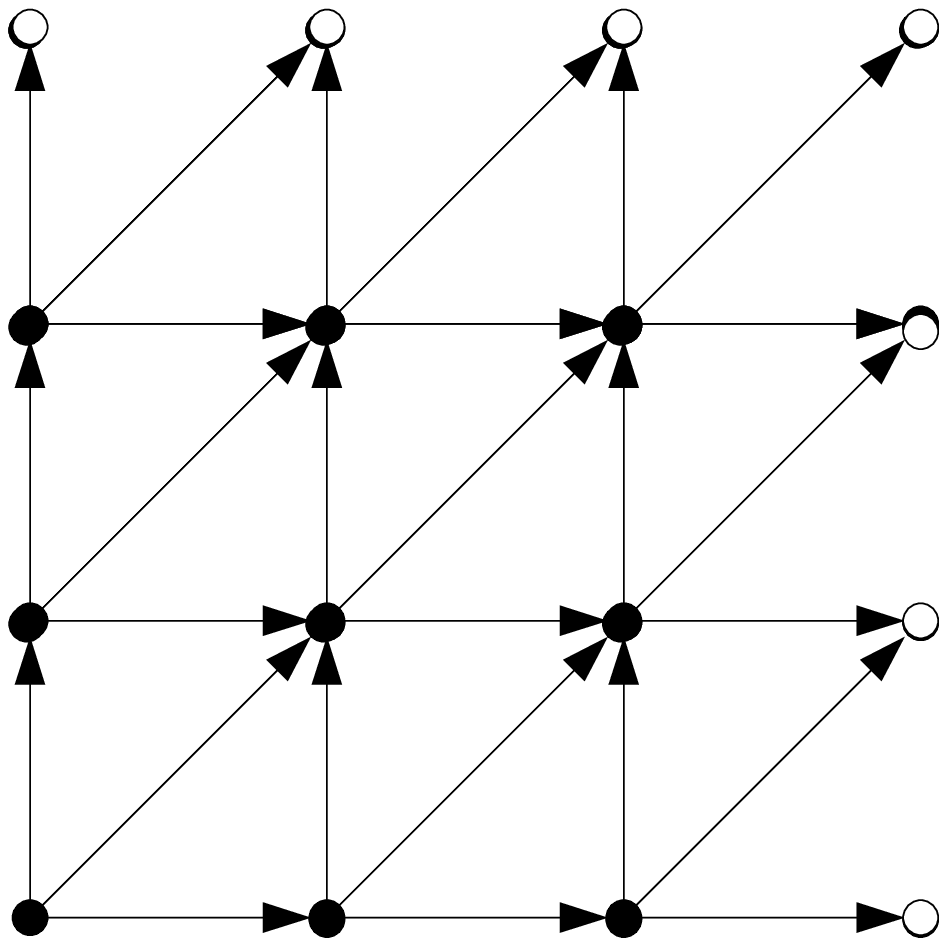}}
\smallskip
\centerline{Figure 2: $\IC^3/\ZN2$ Moose for Bifundamental
Scalars. Note that it is periodic.}
\smallskip

In $4d$ ${\cal N}=1$ language, $3N^2$ bifundamental chiral
superfields survive, plus adjoints from vector
multiplets for each of the $N^2$ $U(k)$ gauge groups.
(Of course, if $p<3$, we have to dimensionally reduce these fields.)

Giving identical vevs to all `horizontal' and `vertical'
bifundamentals in the quiver diagram generates two discretized
dimensions forming a rectangular torus. Giving a further vev to
the `diagonal' bifundamentals gives instead a slanted torus. In
both cases, the resulting low energy theory can be thought of as a
latticized version of the worldvolume theory of $k$ wrapped
D($p+2$)-branes. (See \cite{DeconstructingLS} for details).

\section{Discrete torsion and fuzzy moose}

As we have seen, D$p$-branes on
$\ZN2$ orbifolds deconstruct D$(p+2)$-branes.
We will argue that discrete
torsion makes the two new world-volume coordinates
non-commutative in an appropriate continuum limit.  We begin with a review of
quiver theories on D-branes probing orbifolds with discrete torsion, discuss
the basic strategy, and proceed with an explicit example.

First let us review the physical meaning of discrete torsion,
relate it to a T-dual $B$-field, and give an overview of the
logical structure of our construction.

\subsection{Discrete Torsion as a T-dual $B$-field}

Orbifolds with discrete torsion \cite{DiscreteTorsion} generalize
geometric orbifolds by adding to the twisted sectors of the path
integral phases which depend on the orbifold group elements
defining that sector.  Modular invariance forces the discrete
torsion to lie
in $H^2(\Gamma, U(1))$, i.e. the torsion is a
two-cocycle of the orbifold group.

A trivial example is the torus, $T^2 = \IR^2 / \IZ\times\IZ$, whose partition
function is a sum over winding $\equiv$ twisted sectors
\begin{equation}
\square = \sum_{(a,b\;|\;a',b')}\;\square_{(a,b\;|\;a',b')}.
\end{equation}
Adding discrete torsion ammounts to adding phases of the form
\begin{equation}
\square = \sum_{(a,b\;|\;a',b')}\;\square_{(a,b\;|\;a',b')} 
\;\;e^{2\pi\!i (a\cdot b' - b\cdot a')/n},
\end{equation}
Importantly, this is identical to the partition
function for the torus with a constant background longitudinal $B$-field with
$b=1/n \in (0,1)$, so the torus with discrete torsion is identical
to the torus with constant background $B$-field.  This fits with the
fact that the $B$-field takes values in $H^2(T^2, U(1))$.

Let's consider the $\IC^3/\ZN2$ orbifold with discrete torsion
$\epsilon \in H^2(\ZN2, U(1))=\IZ_N$ labeled by the integer $m'$
(mod $N$) in $\epsilon=e^{2\pi\!i m'/N} \equiv e^{2\pi\!i /n}.$
For $N$ large, this is a sharp cone over base $S^5/\ZN2$, which 
in particular contains a $T^2$ factor.  By taking
$N\to\infty$ (and $\lim\limits_{N\to\infty} m'/N$ fixed) while moving away from the
orbifold fixed point so as to keep the volume of the torus fixed,
we recover the background $\IR^4 \times T^2$, where the
twisted sectors of the orbifold become the 
winding sectors on the torus, so the discrete
torsion of the orbifold becomes discrete torsion on the torus.

Now probe the $\IC^3/\ZN2$ torsion orbifold with a D$0$-brane far
from the fixed point and again consider the constant volume, large
$N$ limit.  From the above, this limit is identical to the theory
of a D$0$-brane on the same torus in the presence of a background
$B$-field.  T-dualizing both legs of the torus gives a D$2$-brane
wrapping the dual torus with a rescaled background $B$-field,
whose worldvolume theory is SYM on a noncommutative torus with
noncommutativity given by $1/b$.  Thus we can realize
noncommutative SYM as the large-$N$ limit of the quiver theory of
a D-brane on a $\IC^3/\ZN2$ orbifold with discrete torsion.

Detailed study of this quiver theory reveals that, in this limit,
the theory becomes precisely SYM on an $N^2$-point fuzzy torus,
with noncommutativity given in terms of the volume of the torus
and the discrete torsion.  But the large $N$ limit of this fuzzy SYM
is exactly SYM on a noncommutative torus.  
Thus the quiver
theory of a D$p$-brane probing a $\IC^3/\ZN2$ with discrete
torsion precisely deconstructs the worldvolume theory of a
D$(p+2)$-brane wrapping a noncommutative torus. 
This can be expressed in the commuting diagram (figure 3) \vskip 3mm

\smallskip
\centerline{\epsfxsize=0.8\textwidth \epsfbox{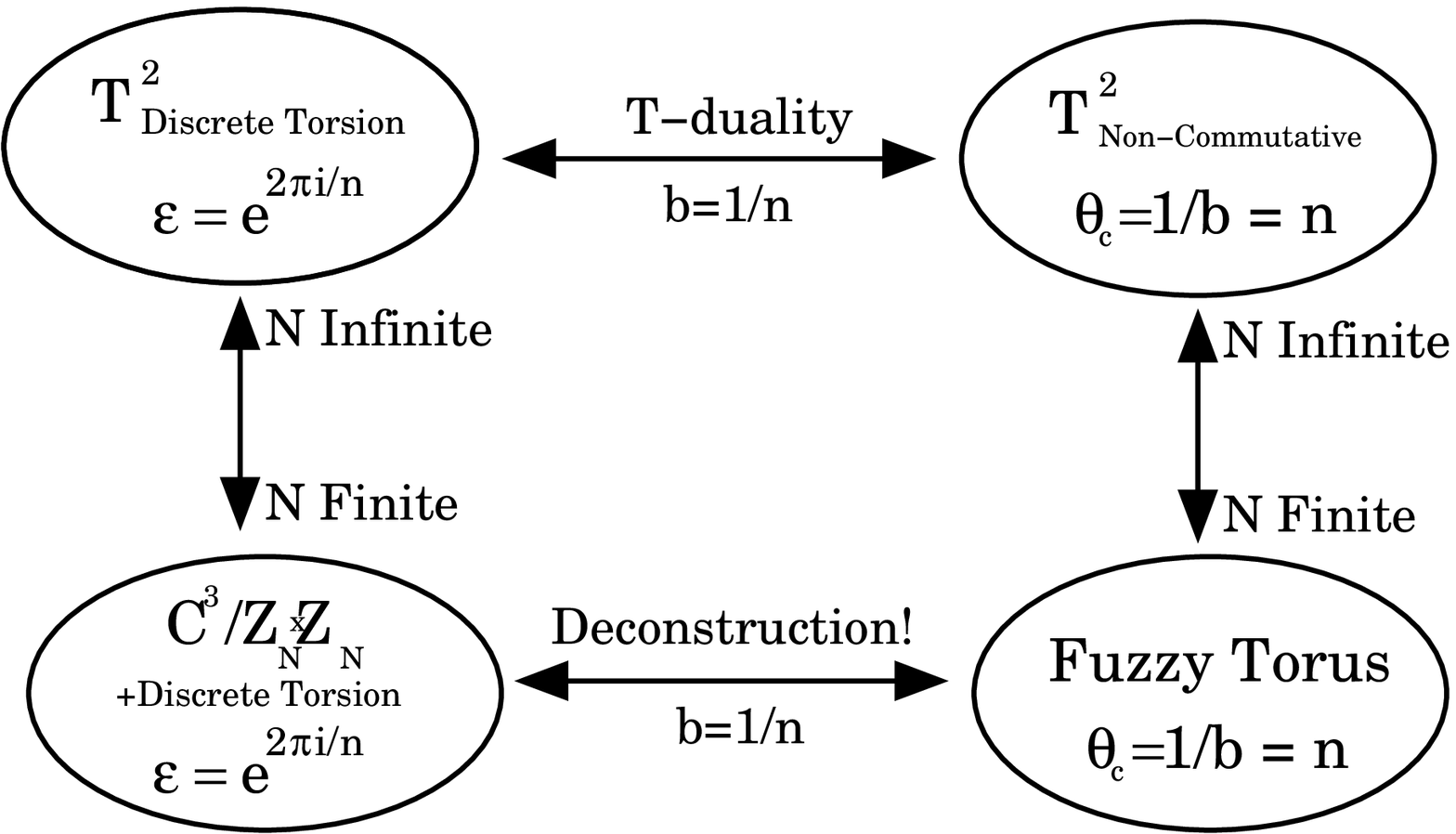}}
\smallskip
\centerline{Figure 3: Deconstruction is T-Duality. Note that
$n\equiv m'/N$.} \centerline{ The scalings indicated are for fixed
volume of the torus seen by closed strings. }
\smallskip

\vskip 5mm

This provides a simple and useful physical interpretation 
of discrete torsion\footnote{We would partiuclarly like to thank
E. Silverstein and K. Dasgupta for discussions on these points.}.
Since the possibility of including discrete
torsion depends on the non-vanishing of $H_2(\Gamma,U(1))$, having 
discrete torsion means that the orbifold can be presented 
as a fibration whose fibres contain 2-cycles. 
The interpretation suggested by our analysis is 
that the discrete torsion should be understood as 
the $B$-field along (the T-dual of) each fibre.
The utility of this interpretation is that it applies just as well in the case
of orbifolds with fixed points (at which points the fibration becomes singular), 
where fractional branes can wrap the shrunken cycle, 
as to orbifolds with freely acting orbifold groups 
(non-singular fibrations), such as $T^2 \equiv \IR^2/\IZ\times\IZ$, where 
there is no shrunken cycle to wrap.

In the remainder of this section we explicitly verify the above
story for $\IC^3/\ZN2$.

\subsection{What moose know about discrete torsion}

As explained in \cite{DiscreteTorsionD-Branes,DiscreteTorsionD-Branes2,Feng2000},
chan-paton indices in the worldvolume theories of D-branes
probing orbifolds with discrete torsion transform
in projective representations of the orbifold group,
$\gamma(g)\gamma(g')=\tilde \epsilon(g,g')\gamma(gg')$,
where the phase $\tilde\epsilon$ again lies
in $H^2(\Gamma, U(1))$.
In the following we focus on the orbifold group $\Gamma =\IZ_N \times \IZ_N$,
for which $H^2(\IZ_N \times \IZ_N, U(1)) \cong \IZ_N$,
so the choice of discrete torsion is specified by one number, $m$.
For $N$ and $m$ relatively prime (which we will assume throughout for simplicity),
there is a unique irreducible
projective representation of the orbifold group, which can be
realized as $N \times N$ matrices involving the usual clock and
shift operators. To get a representation describing $k$
D$p$-branes on the orbifold, we tensor this irreducible
representation with $kN \times kN$ matrices.

%

We emphasize that this phenomenon works for general $B$-fields.
To motivate this, consider deconstruction of a general manifold ${\cal M}$ via
an orbifold with quantum symmetry $\Gamma$ such that $\Gamma \to {\cal M}$ in the
continuum limit.  The discrete torsion is classified by $H^2(\Gamma,U(1))$; in the continuum
limit this becomes $H^2({\cal M},U(1))$, which classifies the $B$-field
along the emergent dimensions.


%
%
%

\subsection{Fuzzy D-branes from $\ZNZN$ orbifolds with discrete torsion}
\label{FuzzyDBranesFromZNZN}
The physics of D-branes in the supersymmetric $\ZNZN$ orbifold with discrete torsion was
described in \cite{DiscreteTorsionD-Branes2}.
The two-cocycle classes $\tilde \epsilon^{\ \! m}$ of $H^2(\Gamma,U(1))\cong
\IZ_N$ will be represented by
\begin{equation}
\begin{array}{c}
 (\ZN2) \times (\ZN2) \ra U(1) \\
((a,b),(a',b')) \ra \zeta^{m(ab' - a'b)},
\end{array}
 \end{equation}
where $\zeta=e^{\pi i/N}$ for $N$ even and $\zeta=e^{2\pi i/N}$
for $N$ odd, $m \in \{0,1\ldots N-1\}$
labels the possible choices of discrete torsion,
and we restrict for simplicity to $N$ and $m$ relatively prime
(as in \cite{DiscreteTorsionD-Branes2}).
With these conventions, $\epsilon \equiv \zeta^{2m}$
generates $\IZ_N$. The geometric action of the orbifold group is
\begin{equation}
  R(e_1) = \epsilon^{- J_{67} + J_{89}}, \quad R(e_2) = \epsilon^{  J_{45} - J_{89}} .
\label{C3ZNZN-RotationsWithDiscreteTorsion} \end{equation}
 We will be interested in the case of $k$ D$(p\leq 3)$-branes,
consisting of $kN^2$ fractional D$p$-branes, probing the orbifold.
In the parent $U(kN^2)$ SYM theory, we have, in the ${\cal N}=1,
d=4$ language, one vector multiplet $\hat A$ and three chiral
superfields $\hat \Phi_1, \hat \Phi_2$ and $\hat \Phi_3$. (In
general, we will use a hat to denote matrices $kN^2 \times kN^2$).

The $\hat \gamma$-matrices acting on the chan-paton sector can be
chosen as
%
%
\begin{equation} \hat\gamma(e_1) = \I1_{k \times k} \otimes \I1_{N \times N}
\otimes U, \quad \hat \gamma(e_2) = \I1_{k \times k} \otimes
\I1_{N \times N} \otimes V .\end{equation}
We define shift and clock matrices $U$ and $V$, satisfying $UV=\epsilon
VU$, as in \cite{DiscreteTorsionD-Branes2} (where
they were called $P$ and $Q$ ). For odd $N$,
\begin{equation} U=\begin{pmatrix}
 0 &1 &0 &\dots &0 \cr
                        0 &0 &1 &\dots &0 \cr
                       \dots &\dots &\dots &\dots &\dots \cr
                        0 &0  &\dots &0 &1 \cr
                        1 &0  &0  &\dots &0
\end{pmatrix} \quad
V=\begin{pmatrix} 0 &\epsilon &0 &\dots &0 \cr
                        0 &0 &\epsilon ^2 &\dots &0 \cr
                       \dots &\dots &\dots &\dots &\dots \cr
                        0 &0  &\dots &0 &\epsilon ^{n-1} \cr
                        1 &0  &0  &\dots &0
\end{pmatrix}.
\end{equation}
For even $N$, $U$ is as above and $V$ is defined using  $\delta^2=\epsilon$ as
\begin{equation} V=\begin{pmatrix} 0 &\delta &0 &\dots &0 \cr
                        0 &0 &\delta ^3 &\dots &0 \cr
                       \dots &\dots &\dots &\dots &\dots \cr
                        0 &0  &\dots &0 &\delta ^{2n-3} \cr
                        \delta ^{2n-1} &0  &0  &\dots &0
\end{pmatrix}
\end{equation}
The fields left invariant by the full orbifold action are of the
form
\begin{equation} \hat A = A \otimes \I1, \quad \hat \Phi_1 = \Phi_1 \otimes U,
\quad \hat \Phi_2 = \Phi_2 \otimes V, \quad \hat \Phi_3 = \Phi_3
\otimes (VU)^{-1}. \end{equation} The superpotential of the parent theory may
be written
\begin{equation} W =  \frac{\tau_p(g_s)}{4(2\pi l_s^2)^2} \hat {\rm Tr} (\hat
\Phi_1 \hat \Phi_2 \hat \Phi_3 - \hat \Phi_1 \hat \Phi_3 \hat
\Phi_2), \end{equation}
where $\tau_p(g_s)$ is the D$p$-brane tension for string coupling
$g_s$. The orbifold-projected superpotential is then
\begin{equation} W =  \frac{\tau_p(g_s)}{4N(2\pi l_s^2)^2} {\rm Tr} ( \Phi_1
\Phi_2 \Phi_3 - \epsilon^{-1}\Phi_1 \Phi_3 \Phi_2),
\label{OrbifoldSuperpotential} \end{equation}
where we have added a factor of $1/|\Gamma|$ as in Section
\ref{OrdinaryD-BraneDeconstruction}. The F-
and D-terms are
\begin{equation} V_F = \frac{\tau_p(g_s)}{4N(2\pi l_s^2)^2} \left( |\Phi_1
\Phi_2 - \epsilon^{-1}\Phi_2 \Phi_1|^2 + |\Phi_2 \Phi_3 -
\epsilon^{-1}\Phi_3 \Phi_2|^2 + |\Phi_3 \Phi_1 -
\epsilon^{-1}\Phi_1 \Phi_3|^2 \right) , \label{F-Term}\end{equation}
\begin{equation} V_D = \frac{\tau_p(g_s)}{16N(2\pi l_s^2)^2} {\rm
Tr}\left([\Phi_1, \Phi_1^\dagger]^2 + [\Phi_2, \Phi_2^\dagger]^2
+[\Phi_3, \Phi_3^\dagger]^2\right), \label{D-Term} \end{equation}
where $|M|^2$ means ${\rm Tr} (M M^{\dagger})$. In the continuum limit,
\begin{equation} \Phi_1= z_1 \I1_{k \times k} \otimes V, \quad \Phi_2= z_2
\I1_{k \times k} \otimes U, \quad \Phi_3= z_3 \I1_{k \times k}
\otimes (VU)^{-1}, \label{BackgroundForFuzzyTorus}\end{equation}
corresponds to a stack of coincident D$(p+2)$-branes; for simplicity
we fix $z_3=0$.
\begin{equation} \Phi_1= z_1 \I1_{k \times k} \otimes V, \quad \Phi_2= z_2
\I1_{k \times k} \otimes U, \quad \Phi_3= 0.
\label{BackgroundForFuzzyTorus-VanishingZ3}\end{equation}
Note that (\ref{BackgroundForFuzzyTorus}) has zero energy and thus
lies on the moduli space.


It is the spectrum of light fluctuations off the moduli space that reveals the
presence of emergent dimensions.  Focussing for clarity on scalars,
we rewrite the bosonic parts $Z_1$
and $Z_2$ of the chiral superfields $\Phi_1$ and $\Phi_2$ as
\begin{equation} Z_1 =(r_1 \I1_{kN \times kN} + H_1) \ \! e^{i \omega_1} L_1 \
\! \TV, \quad Z_2 =(r_2  \I1_{kN \times kN} + H_2) \ \!  e^{i
\omega_2} L_2 \ \!  \TU, \end{equation}
where $H_1, H_2$ are hermitian $kN \times kN$ matrices, $L_1, L_2$
are $kN \times kN$ unitary matrices, and
\begin{equation} z_1 = r_1 e^{i\omega_1}, \quad z_2 = r_2 e^{i\omega_2},\quad
\TU= \I1_{k \times k} \otimes U, \quad \TV= \I1_{k \times k}
\otimes V. \end{equation}
With these conventions, the background
(\ref{BackgroundForFuzzyTorus-VanishingZ3}) corresponds to
$H_1=H_2=0$ and $L_1=L_2=\I1_{kN \times kN}$.

Substituting these into (\ref{F-Term}), we get, among other terms,
\begin{equation} -\ \frac{\tau_p(g_s)r_1^2 r_2^2}{2N (2\pi l_s^2)^2} \
\epsilon^{-1} \ {\rm Tr} \ ( \ L_2 \TU \ \ L_1 \TV \ \
\TU^{\dagger} L_2^{\dagger} \ \ \TV^{\dagger} L_1^{\dagger} \ - \
\I1_{kN \times kN} ) + {\rm c. \! \ c. } \label{Plaquette}\end{equation}
This is precisely the plaquette operator of $U(k)$ gauge theory on
a fuzzy torus \cite{FuzzyTorus}, with $L_1$ and $L_2$ being the
usual link variables! It is a remarkable fact that the full
set of fluctuations flesh out a certain fuzzy torus gauge theory:
besides the $U$-$V$ part (\ref{Plaquette}) of ``$F_{\mu
\nu}F^{\mu\nu}$," we can identify the $p$-$U$ and $p$-$V$ parts
coming form the kinetic terms for $\Phi_1$ and $\Phi_2$. Together
with the original $p$-$p$ piece, this forms the kinetic term  of a
gauge field living in $p$ continuous spacetime dimensions and two
discrete dimensions forming a fuzzy torus.

Further, $H_1$ and $H_2$ appear as adjoint scalars on this fuzzy torus,
the extra pieces of their kinetic terms appearing in (\ref{F-Term}) and
(\ref{D-Term}). For example, the D-term gives ``$({\cal D}_V
H_1)^2$", while ``$({\cal D}_U H_1)^2$" originates from the
F-term.

There are also terms involving other fields and other interactions
which one might not have expected in a simple gauge theory on a
fuzzy torus. While some of them reflect the fact that changing
different expectation values effectively deforms the fuzzy torus
(e.g. from a flat to a slanted torus), some do not have an
immediately obvious interpretation.  This should probably not be
too much of a surprise, as the most naive stringy realization of
gauge theory on a fuzzy torus, i.e. the matrix theory
construction, is unstable.

\subsection{Scalings of the fuzzy moose}
\label{TheComtinuumLimit}

The procedure for taking the continuum limit of fuzzy geometries
is standard and will not be repeated here. Our interest now lies
in finding the scalings of various physical quantities in this
limit. To identify the lattice spacing, we will use the fact that
the normalization of the term in the lagrangian involving
$\dot{L}_1$ and $\dot{L}_2$ is
\begin{equation} \frac{\tau_p(g_s)}{2N} \ \! {\rm Tr}\ \!  \left( \ {r_1^2}
(\dot{L}_1 \TV) (\dot{L}_1 \TV)^{\dagger} + {r_2^2} (\dot{L}_2
\TU) (\dot{L}_2 \TU)^{\dagger}\ \right). \label{L-Dot-Terms} \end{equation}
Comparing (\ref{Plaquette}) and (\ref{L-Dot-Terms}) to the
normalization in \cite{FuzzyTorus} we get
\begin{equation} a^{\TT} = \frac{2\pi l_s^2}{r_2}, \quad a^{\TO} = \frac{2\pi
l_s^2}{r_1}.\end{equation}
The radii and volume of the torus are
\begin{equation} R^{\TT} = \frac{N a^{\TT}}{2 \pi}= \frac{Nl_s^2}{r_2}, \quad
R^{\TO} = \frac{N a^{\TO}}{2 \pi}= \frac{Nl_s^2}{r_1},
\label{FuzzyTorusRadii} \end{equation}
\begin{equation} V_o =  N^2  a^{\TO} a^{\TT} = \frac{(2\pi l_s^2)^2 N^2}{r_1
r_2}. \end{equation}
The most relevant quantity, of course, is the emergent
noncommutativity parameter $\theta^{\ \! \! \tilde 1 \tilde 2}$.
For this purpose, we formally write
\begin{equation} U =e^{ix^{\TT} / R^{\TT}} , \quad V = e^{ix^{\TO} / R^{\TO}},
\label{UV-AsExponencials} \end{equation}
\begin{equation} [x^{\TO}, x^{\TT}] = i \theta^{\ \! \! \tilde 1 \tilde 2}. \end{equation}
Since we defined $\epsilon \equiv \zeta^{2m}$, and $\zeta=e^{\pi
i/N}$ for $N$ even and $\zeta=e^{2\pi i/N}$ for $N$ odd, the
relation $UV=\epsilon VU$ implies
 \begin{equation} \theta^{\ \! \! \tilde 1 \tilde 2} = \frac{2 \pi m}{N} R^{\TO}R^{\TT} \quad (N \ {\rm odd
  }),\quad
 \theta^{\ \! \! \tilde 1 \tilde 2} = \frac{4 \pi m}{N} R^{\TO}R^{\TT} \quad (N \ {\rm even  }). \label{Thetas}\end{equation}
From the overall normalization and using the results of
\cite{FuzzyTorus}, we can also read off the gauge coupling
\begin{equation} G^2_{ym, p+2} = \frac{1}{\tau_p(g_s) l_s^4} \ \!
R^{\TO}R^{\TT}. \end{equation}
We will continue this discussion in section 3.5 after relating the
fuzzy moose to the matrix theory construction of
higher-dimensional D-branes.

\subsection{Large-$N$ matrix theory vs. the giant fuzzy moose}

The standard matrix theory construction of $k$ noncompact
D$(p+2)$-branes with $N\to\infty$ D$p$-branes\cite{MatrixDBranes, Seiberg}
begins with the ordinary matrix
lagrangian in $\IR^{10}$ and expands around a background of
$kN\times kN$ matrices
satisfying
\begin{equation} [x^{\tilde 1}, x^{\tilde 2}] = i \theta^{\ \! \!  \tilde 1
\tilde 2}, \label{CommutatorX1X2} \end{equation}
the other $x^{\tilde i}$ being zero\footnote{For simplicity we
consider generating two noncommutative dimensions.}. The matrices
$x^{\tilde 1}$ and $x^{\tilde 2}$ should generate the space of
$N^2$ linearly independent matrices, so that any $kN\times kN$
matrix can be expressed as a $k \times k$ matrix whose entries are
functions of $x^{\TO}$ and $x^{\TT}$. The background
(\ref{CommutatorX1X2}) satisfies the equations of motion, and by
studying its fluctuations, one recovers the non-commutative field
theory describing the higher-dimensional D-branes.

Note that the matrix potential can be obtained by dimensional
reduction of the kinetic term and the superpotential of ${\cal
N}=4, d=4$ SYM theory. Expressed in the ${\cal N}=1, d=4$
variables, the superpotential is
\begin{equation} W = \frac{\tau_p(g_{mat})}{4(2\pi l_s^2)^2} {\rm Tr} ( \Phi_1
\Phi_2 \Phi_3 - \Phi_1 \Phi_3 \Phi_2),
\label{MatrixSuperpotential} \end{equation}
which reproduces (\ref{OrbifoldSuperpotential}) in the limit
$\epsilon \to 1$, if we choose $g_{mat}=Ng_s$.

A finite $N$ analog of (\ref{CommutatorX1X2}) can be constructed
with the same matrices  as in Sections \ref{FuzzyDBranesFromZNZN}
and \ref{TheComtinuumLimit}, i.e.
\begin{equation} \Phi_1= z_1 \I1_{k \times k} \otimes V, \quad \Phi_2= z_2
\I1_{k \times k} \otimes U, \quad \Phi_3= 0.
\label{FiniteNMatrixBackground} \end{equation}

However, this background does not minimize the potential and
will evolve with time. For this reason, we
should talk only about very large D-branes, for which the decay is
slow. Alternatively, one might add other terms to the potential,
which would stabilize (\ref{FiniteNMatrixBackground}).  To our knowledge,
such a stable construction has not been realized within string theory.

Let's compare this to the construction in Sections
\ref{FuzzyDBranesFromZNZN} and \ref{TheComtinuumLimit}.  The
backgrounds about which we expand,
(\ref{BackgroundForFuzzyTorus-VanishingZ3}) and
(\ref{FiniteNMatrixBackground}), are formally identical, differing
in the $\epsilon^{-1}$ factor in the orbifold
superpotential (\ref{OrbifoldSuperpotential}). In general, this is
an important distinction. On the other hand, we are free to take
$\epsilon \ra 1$ ($m/N \ra 0$). In this limit, the physics of the
two approaches should be the same. Indeed, this can be explicitly checked.
In both cases we end up with a stack of D$(p+2)$-branes. Since we
want to keep $\theta^{\ \! \! \tilde 1 \tilde 2}$ in
(\ref{Thetas}) fixed, the D$(p+2)$-branes will be very large, decompactifying
in the strict limit. This is precisely the situation in
which the matrix theory configuration becomes stable.

\subsection{Fuzzy math and Morita equivalence}

Now, we would like to compare the scalings of various parameters
in the matrix theory (for very large $D(p+2)$-branes) to the
quiver theory scalings found in Section \ref{TheComtinuumLimit}.
At first sight, they seem manifestly different; the matrix
background (\ref{FiniteNMatrixBackground}) describes a torus with
radii $r_1 = |z_1|$ and $r_2 =|z_2|$, while the fuzzy moose radii
(\ref{FuzzyTorusRadii}) are inversely proportional to $r_1$ and
$r_2$!

Sober second thoughts reveal that it is incorrect to compare
the radii in this way. The radii $r_1$ and $r_2$ are as measured
by the closed string metric, while those in
(\ref{FuzzyTorusRadii}) should be compared to open string
quantities.  More precisely, recall that there is
an infinite number of possible descriptions of non-commutative
theories \cite{SeibergWitten, Seiberg}, differing by the choice of
the two-form $\Phi'_{\tilde i \tilde j}$ (not to be confused with
the chiral superfields $\Phi$) appearing in the commutation
relations
\begin{equation} [x^{\tilde i}, x^{\tilde j}] = i \theta^{\ \! \! \tilde i
\tilde j}, \quad
 [\partial_{\tilde i}, x^{\tilde j}] = i \delta_{\tilde i}^{\tilde
 j}, \quad
[\partial_{\tilde i}, \partial_{\tilde j}] = -i \Phi'_{\tilde i
\tilde j} \quad (\tilde i, \tilde j = \tilde 1, \tilde 2)\end{equation}
As explained in \cite{Seiberg}, the choice which matrix theory
naturally selects is
\begin{equation} \Phi' = - B. \label{BAndPhi} \end{equation}
(This applies also to the fuzzy moose theory.) For this value
of $\Phi'$, the relation between open and closed string parameters is
\begin{equation}
\begin{array}{c}
 \theta = B^{-1}, \quad
 G = - (2 \pi l_s^2)^2 B g^{-1}B, \quad
 G_s = g_s \det (2\pi l_s^2 B g^{-1})^{\frac{1}{2}}.
 \label{OpenVsClosed}
\end{array}
 \end{equation}
$G$ and $G_s$ are the open string metric and coupling,
respectively, while $g$ and $g_s$ denote their closed string
counterparts. Here we condense notation, manipulating matrices as
if they had only indices $\tilde 1$ and $\tilde 2$ and suppressing
other components. Using the  continuum results of \cite{Seiberg} in a frame where
\begin{equation} g_{\tilde i \tilde j} = \eta_{\tilde i \tilde j}, \quad
x^{\tilde 1} \in (0,2\pi r_1), \quad x^{\tilde 2} \in (0,2\pi r_2)
\end{equation}
we can express the $B$-field along the brane as \cite{Seiberg}
\begin{equation} B_{\tilde 1 \tilde 2} = \frac{2\pi N}{V_c} = \frac{N}{2 \pi
r_1 r_2}, \end{equation}
where
\begin{equation} V_c= (2\pi)^2 r_1 r_2 \end{equation}
is the volume of the torus as seen
by closed strings.
The open string metric (from \ref{OpenVsClosed}) is
\begin{equation} G_{\tilde i \tilde j} = (2\pi l_s^2)^2\frac{N^2}{(2\pi r_1 r_2
)^2} \eta_{\tilde i \tilde j} \end{equation}
giving the volume seen by open strings
\begin{equation} V_o =  \frac{(2\pi l_s^2)^2 N^2}{r_1
r_2}. \end{equation}
The corresponding gauge coupling is
\begin{equation} G_{ym,p+2} = \frac{1}{\tau_p(g_s)} \ \!  \frac{N^2}{r_1 r_2} ,
\end{equation}
where we have used $g_{mat} = N g_s$, as identified in the
previous section.

 These are exactly the results given by the fuzzy moose,
provided we set $m=1$ in choosing the discrete torsion.  At first
sight this is somewhat disconcerting; why does the moose have this
extra parameter that does not appear in the strictly infinite-$N$
matrix theory, and what does it mean physically, anyway?  Are the
fuzzy moose with different $m$ really different theories?

The resolution comes from morita equivalence in the
non-commutative theory, which derives from T-duality in the
original theory.  As promulgated by Seiberg and Witten
\cite{SeibergWitten}, morita equivalence relates a noncommutative
theory on a flat torus with metric $G$, gauge coupling $g_{ym}$
and rational theta parameter $\Theta = \frac{m}{N}$ to a
commutative theory with parameters
\begin{equation} \Theta' = 0, \quad G'=\frac{G}{N^2}, \quad g_{ym}'=g_{ym}N^{1/2}   \end{equation}
As the rescalings of the metric and gauge coupling do not depend
on $m$, the fuzzy moose with the different $m$ we consider are all
morita equivalent to the same commutative theory on the same torus
with the same coupling, and thus equivalent to each other.


This equivalence must be read with a bit of care.
For finite $N$, T-duality on the orbifold is more subtle than
on the cylinder (in particular, since winding is conserved only $mod$ $N$, the
dual momenum is conserved only $mod$ $N$), so the morita equivalence may be only
approximate.  This in fact seems necessary, since the orbifold theories with discrete
torsion for any finite $N$ and different $m$ (again all relatively prime) appear
manifestly different - the surviving $\IZ_N$ quantum symmetry groups are
embedded differently in the $\ZN2$, and the quiver theory superpotentials
contain different phases.  It is only in the strict large $N$ limit that this
naive T-duality is exact and the theories become truly identical.
Happily, it is precisely in this limit that we compare to the
strictly infinite-$N$ matrix theory construction \cite{Seiberg},
which has no such parameter in the first place.  This remarkable
agreement provides further evidence that the fuzzy moose agrees
with matrix theory in the large $N$ limit. That they disagree
slightly at finite $N$ again should not be worrying: for finite
$N$, the matrix theory background does not solve the F-term
constraints and is thus not stable, while the moose theory is.
Physically, the two presentations are essentially two different
regularizations of the noncommutative theory which need agree only
in the deregulated limit - which they do.

\section{Conclusion and Open Problems}

We have presented considerable evidence that quiver theories
living on D-branes probing orbifolds with discrete torsion
deconstruct higher-dimensional non-commutative theories in the
large-moose limit. The lagrangian of the moose theory far along
its higgs branch reproduces, for large moose, the lagrangian
(\ref{F-Term}), (\ref{Plaquette}) for the fuzzy torus. In the
strict large moose limit this becomes a gauge theory on a
noncomutative torus. The fuzzy moose agrees with the matrix theory
construction in the strict large $N$ limit.

It is remarkable that the fuzzy moose is completely well defined
for all $N$, providing a novel and consistent regularization of
noncommutative theories and an explicit realization in string
theory of gauge theory on a fuzzy torus.  This begs the question
of how the fuzzy moose encodes the UV/IR correspondence of the
continuum non-commutative theory. For example, non-SUSY fuzzy
moose theories should deconstruct non-SUSY noncommutative
theories, in which we expect IR poles in physical processes
arising from UV degrees of freedom \cite{IRSingularities}; how are
these divergences regulated in the fuzzy moose theory?  This
should be a fruitful ground for exploration.

One immediate extension of our construction is to note that since
the strong coupling limit of $4d~{\cal N}=4$ SYM is NCOS theory,
taking the strong coupling limit of the fuzzy moose D$1$-branes
should provide an explicit deconstruction of NCOS theory. It is
also tempting to speculate about theories one might (de)construct
from more baroque orbifold geometries.  Along these lines, it seems
that similar arguments might be used in orbifolds of the conifold with
discrete torsion, which have been studied extensively in \cite{Dasgupta:2000hn}.

Additionally, while the fuzzy moose on D$2$- or D$3$-branes naively
deconstructs noncommutative $D4$- and $D5$-brane theories, arguments similar
to those in \cite{DeconstructingLS} suggest that they should actually deconstruct
some UV completion of these theories, namely some generalization of
$(2,0)$ and little string theories,
or perhaps even some more general 5-brane theory with a 3-form generalization of
non-commutativity.


\section{Acknowledgements}

We would like to thank N.~Arkani-Hamed, A.~Cohen, K. Dasgupta, L.~Motl,
M.~Van~Raamsdonk, E.~Silverstein, L.~Susskind and Y.~Zunger for
very useful discussions and comments on drafts of this paper.
M.~F. would particularly like to thank L.~Motl for extensive
correspondence. A.~A. is supported in part by an NSF Graduate
Fellowship, by DOE contract DE-AC03-76SF00515, and by viewers like
you. M.~F. is supported in part by a Stanford Graduate Fellowship,
by NSF grant PHY-9870115, and by the Institute of Physics, Academy
of Sciences of the Czech Republic under grant no. 
GA-AV{\v C}R~A10100711. 
Both authors are supported by the beautiful planet Earth.

\end{document}